\documentclass[%
reprint,
twocolumns,
aps,
pre,
showpacs,
notitlepage,
nobibnotes,
superscriptaddress
]{revtex4-2}

\usepackage{float}
\usepackage{bigints}
\usepackage{psfrag}
\usepackage{grffile}
\usepackage{verbatim}
\usepackage{microtype}
\usepackage{multirow}
\usepackage{enumitem}
\usepackage{amsmath}
\usepackage{amssymb}
\usepackage{amsthm}
\usepackage{mathrsfs}
\usepackage{mathtools}
\usepackage{graphicx}
\usepackage{tikz}
\usepackage{bbm}
\usepackage{subfig}
\usepackage[linesnumbered,ruled]{algorithm2e}
\usepackage{color}
\usepackage{tcolorbox}

\usepackage{ragged2e}

\definecolor{myblue}{rgb}{0.153,0.322,0.706}
\usepackage[colorlinks,linkcolor=myblue,urlcolor=myblue,citecolor=myblue]{hyperref}
\usepackage{geometry}
\usepackage{url}
\usepackage{hyperref}
\usepackage{xcolor}

\newcommand{\be}{\begin{equation}}
\newcommand{\ee}{\end{equation}}

\def\bc{\begin{center}}
\def\ec{\end{center}}
\def\bea{\begin{eqnarray}}
\def\eea{\end{eqnarray}}

\DeclareMathOperator{\arcsinh}{arcsinh}
\DeclareCaptionJustification{justified}{\justifying}
\definecolor{airforceblue}{rgb}{0.36, 0.54, 0.66}
\definecolor{brickred}{rgb}{0.8, 0.25, 0.33}
\definecolor{amber}{rgb}{1.0, 0.75, 0.0}
\definecolor{applegreen}{rgb}{0.55, 0.71, 0.0}
\definecolor{magenta}{rgb}{0.965, 0, 0.859}

\newcommand{\newtext}[1]{{\color{black}{#1}}}
\newcommand{\newnewtext}[1]{{\color{black}{#1}}}
\newcommand{\deletetext}[1]{}

\newcommand{\ssp}{\nobreak\hspace{0.1em}}
\newcommand{\er}[1]{Eq.\ssp\eqref{#1}}
\newcommand{\ers}[2]{Eqs.\ssp(\ref{#1}-\ref{#2})}
\newcommand{\era}[2]{Eqs.\ssp(\ref{#1}) and (\ref{#2})}

\newcommand{\Er}[1]{Equation\ssp\eqref{#1}}
\newcommand{\Ers}[2]{Equations\ssp(\ref{#1}-\ref{#2})}

\renewcommand{\S}{{\cal S}}
\newcommand{\E}{{\cal E}}
\newcommand{\II}{{\cal I}}
\newcommand{\tII}{\tilde{\cal I}}
\newcommand{\orho}{{\overline{\rho}}}
\newcommand{\oM}{{\overline{M}}}
\newcommand{\oj}{{\overline{j}}}

\begin{document}
\title{Level 2.5 large deviations and uncertainty relations \\ for non-Markov self-interacting dynamics}

\author{Francesco Coghi}
\email{francesco.coghi@leicester.ac.uk}
\affiliation{School of Physics and Astronomy, University of Nottingham, Nottingham, NG7 2RD, UK}
\affiliation{Centre for the Mathematics and Theoretical Physics of Quantum Non-Equilibrium Systems,
University of Nottingham, Nottingham, NG7 2RD, UK}

\author{Amarjit Budhiraja}
\affiliation{Department of Statistics and Operations Research, University of North Carolina, Chapel Hill, NC 27599, USA}

\author{Juan P.\ Garrahan}
\affiliation{School of Physics and Astronomy, University of Nottingham, Nottingham, NG7 2RD, UK}
\affiliation{Centre for the Mathematics and Theoretical Physics of Quantum Non-Equilibrium Systems,
University of Nottingham, Nottingham, NG7 2RD, UK}

\date{\today}

\begin{abstract}
    We address the general problem of formulating the dynamical large deviations of non-Markovian systems in a closed form. 
    Specifically, we consider a broad class of ``self-interacting'' jump processes whose dynamics depends on the past through a functional of a state-dependent empirical observable. 
    Exploiting a natural separation of timescales, \newnewtext{and under the corresponding mixing conditions,} we obtain the exact (so-called ``level 2.5'') large deviation joint statistics of the empirical measure over configurations and of the empirical flux of transitions. 
    As an application of this general framework, we derive explicit general bounds on the fluctuations of trajectory observables, generalising to the non-Markovian case both thermodynamic and kinetic uncertainty relations. 
    We illustrate our theory with simple examples, and discuss potential applications of these results. 
\end{abstract}

\maketitle

\paragraph*{\bf \em Introduction.} Communicating information related to past experiences is, arguably, one of the most peculiar features of intelligent beings. Millions of years of evolution have led small organisms, such as \textit{E.\ Coli}, \textit{P.\ Aeruginosa} and \textit{Pharaoh's ants}, to develop \textit{external spatial memories}: as these organisms navigate space, they create a memory in the surrounding environment by dropping chemicals~\cite{budrene1995dynamics,brenner1998physical,sumpter2003from,jackson2006longevity,zhao2013psl-trails,gelimson2016multicellular} or by inducing structural changes~\cite{reid2012slime}. This memory then serves other micro-organisms of the same \textit{autochemotactic} species to inform future behaviour. Such natural behaviours can also inform strategies~\cite{howse2007self-motile} used to build active self-propelling and interacting artificial agents~\cite{thutupalli2011swarming,hokmabad2022chemotactic,nakayama2023tunable}.

The typical behaviour of autochemotactic organisms can be explained using non-Markov stochastic systems, such as field-based continuum models~\cite{keller1970initiation,tsori2004self-trapping,gelimson2015collective,grafke2017spatiotemporal}, also in combination with stochastic dynamics~\cite{grima2005strong-coupling,sengupta2009dynamics,pohl2014dynamic,kranz2016effective}, and reinforced random processes~\cite{coppersmith1986random,toth2001self-interacting,othmer2006aggregation,pemantle2007a-survey,erschler2011stuck,kious2016stuck,barbier-chebbah2022self-interacting,bremont2024exact}. Self-interacting processes (SIPs)---including discrete-time chains~\cite{moral2007self-interacting,schreiber2001urn-models,coghi2025self-interacting}, continuous-time jump processes~\cite{budhiraja2025jump} and diffusions~\cite{benaim2002self-interacting,kurtzmann2010the-ode-method}---describe dynamics mediated by a functional of the ``empirical occupation measure'', a stochastic field that models changes generated by autochemotactic agents in the environment. Such self-interaction generates long-time memory and strong correlations with the past, leading to non-trivial dynamics, including ergodicity breaking~\cite{benaim2002self-interacting,benaim2005self-interacting,coghi2024current}, and modified (potentially accelerated) first-passage dynamics~\cite{aleksian2024self-interacting,coghi2025accelerated}.

Despite the importance of SIPs, their rare events have attracted systematic attention only recently~\cite{franchini2017large,budhiraja2022empirical,coghi2025self-interacting,budhiraja2025large,coghi2025accelerated,budhiraja2025jump,franchini2017large}. While the dynamical large deviations (LDs) \cite{lecomte2007thermodynamic,garrahan2007dynamical,maes2008canonical,touchette2009the-large,bertini2012large,chetrite2015nonequilibrium,garrahan2018aspects,barato2018a-unifying,jack2020ergodicity}
and their corollaries---such as bounds on fluctuations \cite{barato2015thermodynamic,gingrich2016dissipation,garrahan2017simple,terlizzi2018kinetic,horowitz2020thermodynamic}---are now well established for Markovian systems, for non-Markovian dynamics general results are mostly restricted to semi-Markov processes~\cite{maes2009dynamical,garrahan2010thermodynamics,shreshtha2019thermodynamic,carollo2019unraveling,ertel2022operationally,jia2022large,macieszczak2024ultimate,liu2024semi-markov,maier2025a-pedestrians} (see however \cite{falasco2020unifying,di-terlizzi2020a-thermodynamic,koyuk2020thermodynamic}).

Here we address this gap by studying a broad class of non-Markovian processes, specifically self-interacting jump processes (SIJPs), where interaction with the past is mediated by a functional of a general empirical observable. We analytically characterise dynamical fluctuations of SIJPs by deriving their so-called {\em level 2.5 of large deviations}, that is, the joint statistics of the long-time empirical occupation of configurations and empirical flux of transitions between them. From this fundamental result, and in analogy to what occurs in the Markovian case \cite{gingrich2016dissipation,garrahan2017simple,macieszczak2024ultimate}, one can derive general fluctuation bounds for dynamical observables, which allows us to extend both thermodynamic \cite{barato2015thermodynamic,gingrich2016dissipation} and kinetic \cite{garrahan2017simple,terlizzi2018kinetic} uncertainty relations (TURs and KURs) to this wide class of non-Markov processes. 

Below we present our general results for the level 2.5 LDs and fluctuation bounds of SIJPs. The key technical advance is the observation at long times of a separation between the relevant timescales for the evolution of empirical measures, which determine the instantaneous transition rates of a SIJP, and those necessary for the microscopic jump dynamics to become fully mixing. Formal proofs appear in our accompanying long paper \cite{coghi2026proof}. We illustrate our theory with two simple examples, and discuss broader implications of our results.

\paragraph*{\bf \em Self-interacting jump processes.} We define a SIJP as the continuous-time chain $\left( X_t \right)_{0 \leq t \leq T}$ of discrete configurations, $x \in \S$ with $|\S| = d$, with $\E = \left\lbrace (x,y) \in \S \times \S: x \neq y \right\rbrace$ denoting the set of all possible jumps between them (for notation and conventions see the \hyperref[appendix]{Appendix}). At time $t$, the SIJP jumps from the current state $x$ to $y \neq x$ according to a rate $Q_{xy}(A_t) \in \mathbb{R}_+$, that depends on a general state-dependent \textit{empirical observable} of the process~\cite{touchette2009the-large,chetrite2015nonequilibrium}
\begin{equation}
    \label{eq:EmpObs}
    A_t = t^{-1} \int_0^t f_{X_{t'}} \, d{t'}  \, 
\end{equation}
with $f: \S \rightarrow \mathbb{R}$ a bounded function. In order for the matrix $Q(A_t)$ to be a stochastic generator, we define its diagonal components to be the negative escape rates 
\begin{equation}
    \label{eq:DiagRateSIJP}
    Q_{xx}(A_t) = - \sum_{\substack{y \in \S,  y \neq x}} Q_{xy}(A_t) \, .
\end{equation}
The dependence of the generator at time $t$ on the trajectory up to that point through \er{eq:EmpObs} makes the process non-Markovian. 

Two central objects of interest are the {\em empirical measure} \cite{touchette2009the-large,bertini2012large}
\begin{equation}
    \label{eq:EmpOccSIJP}
    L_x(t) \coloneqq
    t^{-1} \int_0^t \mathbf{1}_x(X_{t'}) 
    d{t'} 
    \, ,
\end{equation}
viz.\ the fraction of time that the current realisation of the process has spent in each configuration in $\S$, and the \textit{empirical flux} \cite{touchette2009the-large,bertini2012large}
% \begin{equation}
%     \label{eq:EmpFluxSIJP}
%     \Phi_{xy}(t) \coloneqq
%         \lim_{\epsilon \to 0}    
%         \,
%         t^{-1} \int_0^t 
%         \mathbf{1}_{x}(X_{{t'-\epsilon}}) 
%         \mathbf{1}_{y}(X_{t'}) 
%         d{t'} 
%         \, ,
% \end{equation}
\begin{equation}
    \label{eq:EmpFluxSIJP}
     \Phi_{xy}(t) \coloneqq t^{-1} \sum_{t', \Delta X_{t'} \neq 0} \mathbf{1}_{x}(X_{t_{-}'}) 
         \mathbf{1}_{y}(X_{t'}) 
\end{equation}

viz.\ the number of jumps per unit time for every $(x,y) \in \E$.

We assume that $Q(A_t)$ is irreducible, and therefore positively recurrent over $\S$, for all $t \in \mathbb{R}_+$ and $A_t: \mathbb{R}_+ \rightarrow \mathbb{R}$ (generalisation to vector valued $A_t$ is straightforward). 
\newtext{
    Additionally, we require the map $a \mapsto Q(a)$ to be at least $C^1$ for $a \in \mathbb{R}$ and rates $Q_{xy}(A_t) \geq q_- >0$ for all $(x,y) \in \mathcal{E}$. That is, for no value of the argument of $Q(\cdot)$ the transition rates for the directed links in $\E$ can be smaller than some constant $q_-$.      
    } 
We further assume the existence of at least one \newtext{attracting} stationary state, which, unlike in Markov processes for the same assumptions, need not be unique. This implies that the empirical measure and flux converge in the long-time limit to typical values which are not necessarily unique. 
\newtext{These assumptions are discussed in detail in Ref.\ssp\cite{coghi2026proof}.
}

For a Markov process with fixed generator $Q$, the stationary state $\pi$ obeys $\pi Q = 0$, and the empiricals converge to $\pi$ and $\pi \circ Q$ \cite{touchette2009the-large,bertini2012large}. In the SIJP case one needs the long-time limit of the observable, 
\begin{equation}
    A_t \rightarrow
    \overline{a} = \sum_{x \in \mathcal{S}} f_x \pi_x \eqqcolon f \pi \, , 
\end{equation}
and the stationary state is a solution of the (in general non-linear) equation
\begin{equation}
    \label{eq:StatSIJP}
    \pi Q(\overline{a}) = \pi Q(f \pi) = 0 \, .
\end{equation}
For a SIJP the empiricals converge to
\begin{equation}
    \label{eq:TypBehav}
    L(t) \rightarrow \pi 
    \, , \;\;
    \Phi(t) \rightarrow \varphi
    \, ,
\end{equation}
for some pair $(\pi,\varphi)$, which may depend on the trajectory and is not necessarily unique. We denote the set of limiting pairs $(\pi,\varphi)$ as $\mathcal{A}_{\text{stat}}$. 

\paragraph*{\bf \em Large deviations at level 2.5 for SIJPs.} 
We now focus on the joint probability of the empirical  measure \eqref{eq:EmpOccSIJP} and empirical flux \eqref{eq:EmpFluxSIJP}, $P_T(\ell,\phi) \coloneqq P(L(T) = \ell, \Phi(T) = \phi)$, at some long time $T$. We will show below that these dynamical quantities obey a large deviation principle
\begin{equation}
    \label{eq:Level25Joint}
    P_T(\ell,\phi) \asymp e^{-T I_{\text{2.5}}(\ell,\phi)} \, ,
\end{equation}
where $I_{\text{2.5}}(\ell,\phi)$ is the level 2.5 LD {\em rate function} for SIJPs. From this general rate function for the fluctuations of $L_T$ and $\Phi_T$, the LDs of any time-additive observable [like that of \er{eq:EmpObs}, but also those dependent on jumps] follow by contraction. 

The main ingredients for obtaining \er{eq:Level25Joint} are as follows
(see the accompanying Ref.\ \ssp\cite{coghi2026proof} for full details of derivations). The first is a formulation of the path probability of a SIJP. It has the same product structure of transition rates and waiting time factors as in a Markov process, with two distinctions: the transition rates change in time through their dependence on $A_t$, and the waiting times factors are no longer the exponential of a fixed escape rate times time, but of the integral of the time-varying escape rate. Noting these departures from Markovianity, the next step is to exponentially ``tilt'' this path measure, in analogy to what is done in the Markov case. 

The third step follows after making the key observation of a\newtext{n asymptotic} separation of timescales. At a given time $t$ the SIJP has a  characteristic time for relaxation coming from its generator $Q(A_t)$, for example that set by the smallest escape rate. On the other hand, $Q(A_t)$ itself changes over time due its dependence on the empirical observable. However, $A_t$ accumulates $O(1)$ changes over in $O(t)$ times, cf.\ \er{eq:EmpObs}. This means that in the long-time \newtext{(large deviation)} limit the dynamics of the ``instantaneous'' SIJP has enough time to relax before its rates change appreciably due to the change in $A_t$. \newtext{This can be interpreted as an adiabatic relation: fluctuations in the large deviation limit are fundamentally coupled with the dynamical changes of $A_t$, hence $Q(A_t)$, but homogenised over the instantaneous SIJP.}

As a final step, for convenience, we make 
\newtext{ 
    a sequence of
    two changes of variable inside time integrals. First, we introduce the adimensional logarithmic time $s=\log (t / t_0)$, where $t_0$ is interpreted as the system’s relevant memory horizon and is then set to $1$ without loss of generality. As $t \in [0,T]$, $s \in (-\infty, \log{T}]$. Within this time scale,
    } 
the empirical observable $A_t$ makes $O(1)$ changes in $O(1)$ time\newtext{. Then, we make the change $\tau = \log{T} - s$ to measure backwards from the final logarithmic time $\log \, T$. Due to this change, the integral boundaries flip to $[0,\infty)$. Taking the limit $T \rightarrow \infty$, we get} 
\deletetext{
    the change $t \rightarrow e^t$ inside time integrals [so that in this rescaled time the empirical observable makes $O(1)$ changes in $O(1)$ time], and measure backwards from the final time $T$ by making the flip $t \rightarrow T - t$. In doing so, we treat time as dimensionless by rescaling it with an intrinsic timescale $t_0$, which is interpreted as the system’s relevant memory horizon and is then set to $1$ without loss of generality. }
\begin{equation}
    \label{eq:RateFunct25}
    I_{\text{2.5}}(\ell, \phi) 
    = 
    \inf_{(\rho, H) \text{~s.t.~} (\ell,\phi)} \, \II[\rho,H] \, ,
\end{equation}
with
\begin{align}
    \label{eq:FunctionalFinal25}
    & \II[\rho,H] \coloneqq  
    \\
    & 
    \sum_{(x,y) \in \E}
    \int_{0}^\infty e^{-\tau}   
    \rho_x(\tau) Q_{x y} (M_\tau) \, 
    \Gamma \left( \frac{H_{xy}(\tau)}{Q_{x y}(M_\tau)} \right) 
    \, d\tau \, ,
    \nonumber 
\end{align}
Here $\Gamma(x) \coloneqq x \log x - x + 1$ is the Cram\'{e}r (or relative-entropy) rate function for Poisson statistics. In \era{eq:RateFunct25}{eq:FunctionalFinal25} there are three auxiliary quantities: $H(\newtext{\tau})$ is the generator of a time-dependent but Markovian auxiliary dynamics, $\rho(\newtext{\tau})$ is its ``accompanying'' distribution (instantaneous stationary state)
\begin{equation}
    \label{eq:AccDistrReverse}
    \rho(\newtext{\tau}) H(\newtext{\tau}) = 0 
    \, ,
\end{equation}
and $M_\newtext{\tau}$ the corresponding dynamical observable, \er{eq:EmpObs}, if the empirical density is $\rho(\newtext{\tau})$
\begin{equation}
    \label{eq:Ms}
    M_\newtext{\tau} = 
        e^\newtext{\tau} \int_\newtext{\tau}^\infty e^{-\newtext{\tau}'}  
        \sum_{x \in \S} f_x \rho_x(\newtext{\tau}') d{\newtext{\tau}'} 
    \, .
\end{equation}
Furthermore, the minimisation in \er{eq:RateFunct25} is such that 
\begin{align}
    \label{eq:RareDensity}
    \ell_x &= \int_0^\infty e^{-\newtext{\tau}} \rho_x(\newtext{\tau}) \, d{\newtext{\tau}}
    \, , \\
    \label{eq:RareFlux}
    \phi_{xy} &= \int_0^\infty e^{-\newtext{\tau}'} \rho_x(\newtext{\tau}) H_{xy}(\newtext{\tau}) \, d{\newtext{\tau}}
    \, .
\end{align}
\Er{eq:RateFunct25} with definition \eqref{eq:FunctionalFinal25} and constraints \ers{eq:AccDistrReverse}{eq:RareFlux} provide the explicit form of the level 2.5 LDs for SIJPs. This is the central result of this paper. 

The functional $\II[\rho,H]$ is associated to a trajectory of the SIJP of time extent $T \gg 1$.
It is local in $\rho(\newtext{\tau})$ and $H(\newtext{\tau})$ and therefore extensive in $T$\deletetext{(a property that becomes evident in the original time frame)}. The level 2.5 rate function $I_{\text{2.5}}(\ell, \phi)$ 
is obtained from this functional by averaging over the ensemble of all SIJP's trajectories. In the LD asymptotics, this translates into the infimum over $(\rho(\newtext{\tau}),H(\newtext{\tau}))$, constrained on having the target long-time empirical measure and flux, giving rise to constraints \eqref{eq:RareDensity} and \eqref{eq:RareFlux}. 

\Ers{eq:RateFunct25}{eq:FunctionalFinal25} can be interpreted as a dynamical reweighting of the Poissonian relative-entropy density $\Gamma(x)$, thus generalising the level 2.5 LDs for Markov jump processes~\cite{maes2008canonical,bertini2012large,barato2015a-formal}. Unlike the classical Markov case, the contraction over $(\rho,H)$ generically induces non-convexity of the functional, opening the way to richer dynamical behaviour in the generation of fluctuations of SIJPs. \deletetext{However,}\newtext{Although} none of the examples discussed below exhibit this property, \deletetext{and further work is needed to identify suitable cases} \newtext{in the companion paper~\cite{coghi2026proof} we present an example of dynamical bistability which leads to non-convenxity and a change of speed in the large deviations.} Additionally, the factor $e^{-\newtext{\tau}}$ acts as an infinite-horizon temporal discount: contributions at large $\newtext{\tau}$ (corresponding to short times in the original time frame) are exponentially suppressed, reflecting that early fluctuations of the empirical measure carry a smaller cost than long-time deviations.

The minimiser $(\rho_{\rm opt}(\newtext{\tau}), H_{\rm opt}(\newtext{\tau}))$ of \er{eq:RateFunct25}, which is not necessarily unique\deletetext{ due to non-convexity}, represents the optimal way to produce the fluctuation $(\ell, \phi)$. Note that this corresponds to a time-dependent yet Markovian dynamics generated by $H_{\rm opt}(\newtext{\tau})$. For the special case of the SIJP generator $Q(A_\newtext{\tau})$ being independent of $A_\newtext{\tau}$, which makes the SIJP into a standard Markov jump process,\deletetext{it is easy to see that} \ers{eq:RateFunct25}{eq:RareDensity} reduce to the standard Markovian level 2.5 LD result \cite{maes2008canonical,bertini2012large,barato2018a-unifying}: in this case $(\rho_{\rm opt}, H_{\rm opt})$ become time-independent, the integral in \er{eq:FunctionalFinal25} is trivial, and $(\ell, \phi)$ coincide with $(\rho_{\rm opt}, \rho_{\rm opt} \circ H_{\rm opt})$. 
\deletetext{A notable special case is SIJPs with rate matrices that are affine functions of the empirical density $L(t)$, whose large deviations were rigorously analysed in Ref.\ \ssp\cite{budhiraja2025jump}.} \deletetext{\Ers{eq:RateFunct25}{eq:RareDensity} extend these rigorous results beyond affine SIJPs to a broader class. While our results are exact in the statistical-physics sense, they lack the full mathematical rigour of Ref.\ \ssp\cite{budhiraja2025jump}.}

\newtext{As is customary in tilting arguments, we assume the existence of a time-inhomogeneous process that asymptotically realises the target fluctuation while minimising the path-space relative entropy with respect to the original SIJP. The resulting calculation leads to a local, time-extensive rate functional, which provides strong evidence for the validity of the former assumption and corresponding large deviation principle. In Ref.\ssp\cite{budhiraja2025jump}, this programme is made rigorous for affine dependence on $A_t$. Here, \ers{eq:RateFunct25}{eq:RareDensity} extend the formal derivation to bounded functions on $S$, leaving a fully rigorous mathematical proof for future work.}  \deletetext{we aim to close in future work.}

\paragraph*{\bf \em Kinetic uncertainty relation for SIJPs.} 
The general level 2.5 LD result above is useful in practice in two ways. The first is that the long-time LDs of any time-extensive trajectory observable can be obtained by {\em contraction} from the LDs of the empirical measure and flux. The second is that \er{eq:RateFunct25} is variational, so that even if the exact minimisation is difficult, an approximate one can provide a bound to the true rate function. This latter is the approach used to prove TURs and KURs using LD methods \cite{gingrich2016dissipation,garrahan2017simple}.

Consider a generic flux observable, that is, a quantity that depends additively on the jumps in a trajectory. For a given trajectory it can be written as a contraction with the empirical flux 
\begin{equation}
    \label{eq:GenFluxSIJP}
    B_T = \sum_{(x,y) \in \E} \beta_{xy} \Phi_{xy}(T) \, .
\end{equation}
Assuming a unique stationary state, at long times $B_T$ will become typical, $B_T \rightarrow \overline{b} \coloneqq \sum_{\substack{(x,y) \in \E}} \beta_{xy} \varphi_{xy}$ in probability, cf.\ \er{eq:StatSIJP}. The probability of fluctuations will have an LD form, $P(B_T = b) \asymp e^{-T I(b)}$, where the rate function $I(b)$ can be obtained from the level 2.5 one \eqref{eq:RateFunct25} with the extra constraint 
\begin{equation}
    \label{eq:bconst}
    b = \sum_{\substack{(x,y) \in \E}} \beta_{xy} \phi_{xy}
    \, .
\end{equation}

While exact minimisation of \er{eq:RateFunct25} under \er{eq:bconst} is in general not doable, we can get an approximation with a simple Ansatz where the empirical density coincides with the stationary state, $\ell^* = \pi$, but the flux is a rescaling of the stationary one, $\phi^* = (b / \overline{b}) \, \varphi$, cf.\ Refs.\ \ssp\cite{gingrich2016dissipation,garrahan2017simple}. In order to satisfy the \ers{eq:Ms}{eq:RareFlux} we have to set 
\begin{equation}
   \rho^*(\newtext{\tau}) = \orho(\newtext{\tau})
   \, , \;\;
   H^*(\newtext{\tau}) = (b / \overline{b}) \, Q(\oM_\newtext{\tau})
   \, ,
\end{equation}
that is, the time-dependent generator of the auxiliary process is simply the typical generator of the SIJP rescaled by an overall factor, and where the typical $\orho$ is defined via \era{eq:AccDistrReverse}{eq:Ms} for the typical $H(\newtext{\tau}) = Q(\oM_\newtext{\tau})$. 

Inserting into \er{eq:RateFunct25} we obtain the general bound
\begin{equation}
    \label{eq:IbBoundSIJP}
    I(b) \leq 
    k_{\text{SIJP}} \, 
    \Gamma( b / \bar{b}) 
    \, ,
\end{equation}
where 
\begin{equation}
    \label{eq:DynActSIJP}
    k_{\text{SIJP}} \coloneqq 
    \sum_{\substack{(x,y) \in \E}} 
    \int_0^\infty e^{-\newtext{\tau}} \,
    \orho_{x}(\newtext{\tau}) 
    \,
    Q_{xy}(\oM_\newtext{\tau}) \, d\newtext{\tau} 
    \, ,
\end{equation}
is the average \textit{SIJP dynamical activity}. It generalises the standard dynamical activity \cite{lecomte2007thermodynamic,garrahan2007dynamical,maes2020frenesy:} of Markov jump processes to the non-Markovian case of SIJPs which corresponds to the number of configuration changes per unit time. 

\newnewtext{Assuming that the observable \eqref{eq:GenFluxSIJP} satisfies a central limit theorem in the long-time limit, }\deletetext{E}\newnewtext{e}xpanding \er{eq:IbBoundSIJP} to second order around $\overline{b}$ we obtain a bound on the asymptotic variance of $B_T$, 
\begin{equation}
    \label{eq:KURSIJP}
    \frac{{\rm var}(b)}{\overline{b}^2}
    \geq
    \frac{1}{k_{\text{SIJP}}}
    \, ,
\end{equation}
which we refer to as SIJP-KUR, generalising the KUR of time-homogeneous Markov processes~\cite{garrahan2017simple,terlizzi2018kinetic}. As for Markov processes, the average dynamical activity of a SIJP provides an elementary bound to the quadratic precision of observables.

\begin{figure*}[t!]
  \centering
    \includegraphics[width=0.245\textwidth,height=0.125\textheight,keepaspectratio=true]{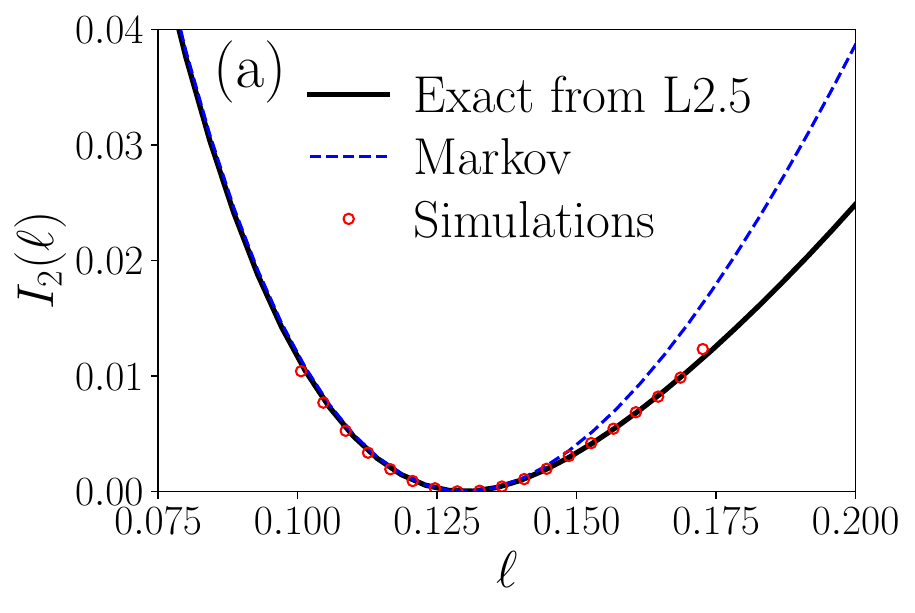} \hfill
    \includegraphics[width=0.245\textwidth,height=0.125\textheight,keepaspectratio=true]{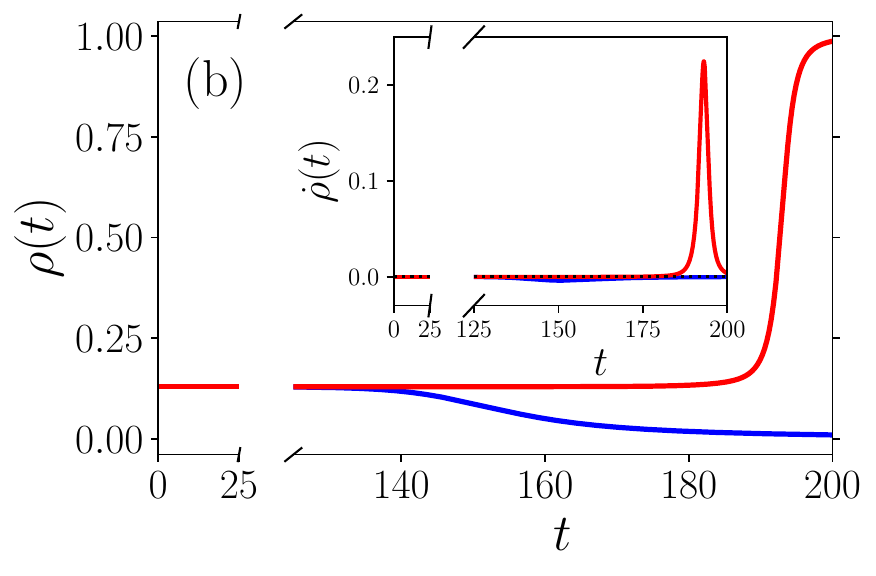} \hfill
    \includegraphics[width=0.245\textwidth,height=0.125\textheight,keepaspectratio=true]{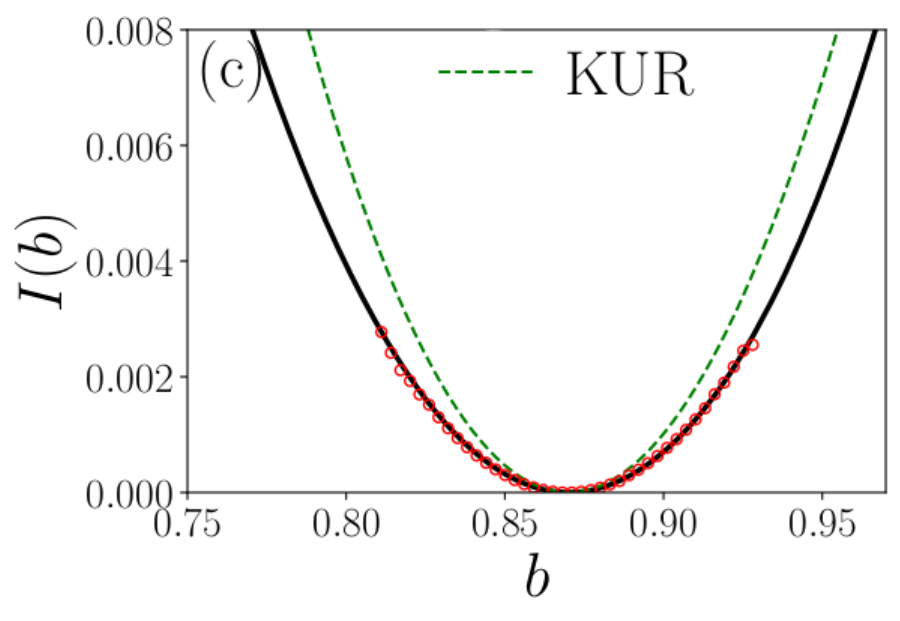}
    \hfill
    \includegraphics[width=0.245\textwidth,height=0.125\textheight,keepaspectratio=true]{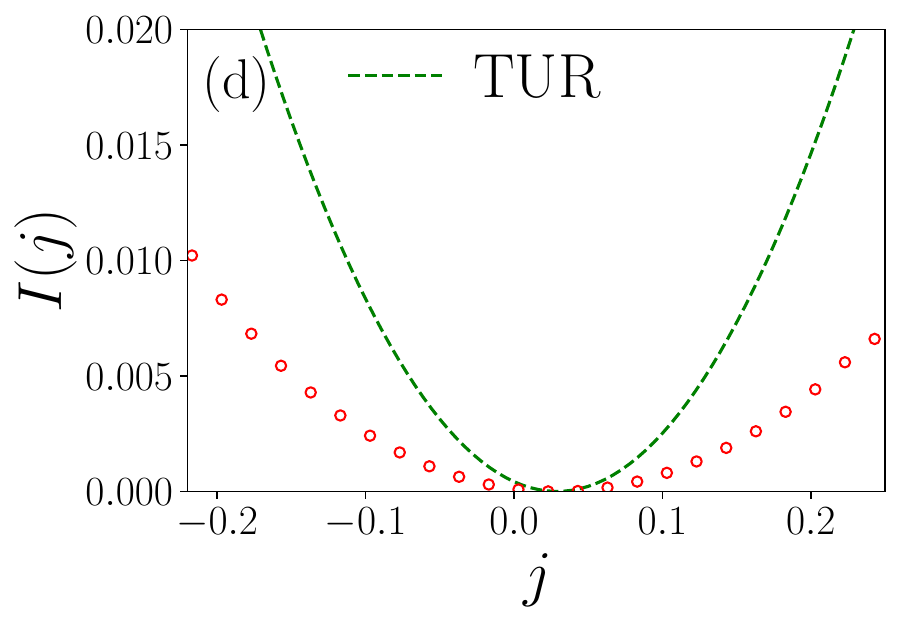}
  \caption{
    (a) Exact rate function for the empirical measure, $I_2(\ell)$ (full-black curve), obtained by solving \era{eq:RateFunct25}{eq:FunctionalFinal25} with constraints \eqref{eq:AccDistrReverse}-\eqref{eq:RareDensity}, for the two-level SIJP with $\alpha = 2$, Monte Carlo simulations (open-red circles), and Markovian approximation (dashed-blue curve).
    (b) Optimal trajectories $\rho(\newtext{\tau})$: 
    \newtext{in backward time $\tau$,} $(\pi_0,\pi_1)$ 
    is an unstable fixed point 
    \newtext{and} the dynamics is integrated 
    \deletetext{backward in time,} 
    \newtext{from $\tau=200$ (which for this example is large enough to be in the asymptotic limit in $\tau$) back to $\tau=0$ for numerical stability.}
    \deletetext{, and} 
    \newtext{B}y progressively truncating the \newtext{red and blue} trajector
    \deletetext{y}
    \newtext{ies} at different 
    \newtext{$\tau=\hat{\tau}$}, one uncovers the most likely dynamics responsible for the fluctuation \deletetext{in} 
    \newtext{$\ell \equiv \rho_{\hat{\tau}}$}. 
    Since trajectories yielding $\ell < \pi_0$ evolve slowly (see Inset), the left tail of $I_2(\ell)$ is close to that of the Markovian approximation (as the exponential discount suppresses distant-in-time contributions faster than they can accumulate). This is not the case for $\ell > \pi_0$, and the non-Markovian character becomes apparent in the higher likelihood of those fluctuations. 
    (c) Exact rate function $I(b)$ for the flux observable $B_T=\Phi_{01}(T)$ (full-black curve) for the two-level SIJP, Monte Carlo simulations (open-red circles), and the SIJP-KUR bound (dashed-green curve).
    (d) Fluctuations of the particle current in the three-state model: Monte Carlo simulations (open-red circles) and the SIJP-TUR bound. 
    All simulation results are for $10^6$ trajectories of time extent $T=1000$.
  }
  \label{fig:all}
\end{figure*}

\paragraph*{\bf \em Thermodynamic uncertainty relation for SIJPs.}
While the SIJP-KUR is valid for all flux-like observables \eqref{eq:GenFluxSIJP}, it can be refined, as in the Markovian case,  for currents, for which $\beta_{xy} = -\beta_{yx}$. This is done by writing the time-dependent flux in terms of its symmetric and antisymmetric parts, $\rho(\newtext{\tau}) \circ H(\newtext{\tau}) \eqqcolon (v(\newtext{\tau}) + u(\newtext{\tau}))/2$, and solving exactly in \era{eq:RateFunct25}{eq:FunctionalFinal25} for the symmetric part $u(\newtext{\tau})$, cf.\ Refs.\ssp\cite{maes2008canonical,bertini2012large,barato2015a-formal}. Doing this, the functional $\II$ in \er{eq:FunctionalFinal25} is replaced by 
\begin{equation}
    \label{eq:CurrRateSimplTURSIJP1}
    \tII[\rho, v] 
    \coloneqq
    \sum_{x < y}
    \int_0^\infty e^{-\newtext{\tau}} 
    \Psi \left( v_{xy}(\newtext{\tau}),v_{xy}^\rho(\newtext{\tau}),u_{xy}^\rho(\newtext{\tau}) \right)
    \, d\newtext{\tau}
    \, ,
\end{equation}
where
\begin{align}
    v_{xy}^\rho(\newtext{\tau}) 
    &\coloneqq 
    \rho_x(\newtext{\tau}) Q_{xy}(M_\newtext{\tau}) - \rho_y(\newtext{\tau}) Q_{yx}(M_\newtext{\tau})
    \, ,
    \\
    u_{xy}^\rho(\newtext{\tau}) 
    & \coloneqq 
    \rho_x(\newtext{\tau}) Q_{xy}(M_\newtext{\tau}) + \rho_y(\newtext{\tau}) Q_{yx}(M_\newtext{\tau})
    \, ,
    \\
    \Psi(p, q, r) 
    &\coloneqq 
    p \arcsinh \frac{p}{\sqrt{r^2-q^2}} 
    - 
    p \arcsinh \frac{q}{\sqrt{r^2-q^2}} 
    \nonumber \\
    &
    - 
    \left( 
        \sqrt{p^2 - q^2 + r^2} - r
    \right) 
    \, ,
\end{align} 
and the minimisation in \er{eq:RateFunct25} is done over $\rho(\newtext{\tau})$ and $v(\newtext{\tau})$, with constraint \eqref{eq:RareFlux} replaced by 
\begin{equation}
    \label{eq:RareFlux2}
    \phi_{xy} - \phi_{yx}
    = 
    \int_0^\infty e^{-\newtext{\tau}} v_{xy}(\newtext{\tau}) \, d{\newtext{\tau}}
    \, .    
\end{equation}

The LD rate function $I(j)$ of any current observable
\begin{equation}
\label{eq:GenCurrSIJP}
    J_T = 
    \sum_{x < y} 
        \beta_{xy} (\Phi_{xy}(T) - \Phi_{yx}(T))
    \, ,
\end{equation}
can be obtained from the level 2.5 \eqref{eq:RateFunct25} by contraction using \er{eq:CurrRateSimplTURSIJP1} with constraints (\ref{eq:AccDistrReverse}-\ref{eq:RareDensity}), and (\ref{eq:RareFlux2},\ref{eq:GenCurrSIJP}). Even if the exact calculation is not practical, one can obtain a general upper bound to $I(j)$. The steps to do this are analogous to those taken to derive the TUR \newtext{using an LD approach} in the Markovian case \cite{gingrich2016dissipation}\newtext{, although more general derivations are available nowadays in that context~\cite{dechant2018entropic,falasco2020unifying}}. 

The first step, after assuming a unique stationary state, is to choose the same Ansatz as in the SIJP-KUR for the empirical density, $\ell^* = \pi$, which in turn implies that $\rho^*(\newtext{\tau}) = \orho(\newtext{\tau})$. The second step is to bound from above the function $\Psi ( v_{xy}, v_{xy}^\orho, u_{xy}^\orho )$ with a form quadratic in $v_{xy}$, as is done in the proof of the TUR \cite{gingrich2016dissipation}. The third step is to perform the minimisation over $v(\newtext{\tau})$ in the resulting bound of the functional $\tII[\orho, v]$. An application of the Cauchy--Schwarz relation eventually yields 
(see Ref.\ \ssp \cite{coghi2026proof} for detailed derivations)
\begin{equation}
    \label{eq:TURBoundRate}
    I(j) 
    \leq 
    \frac{(j - \oj)^2}
    {4 \int_0^\infty e^{-\newtext{\tau}} 
        (j_\newtext{\tau}^{\orho} / \sigma_\newtext{\tau})
    \, dt} 
    \, .
\end{equation}
where 
\begin{align}
    \label{eq:TypTrajCurr}
    j_\newtext{\tau}^{\orho} 
    &\coloneqq 
    \sum_{x < y} 
        \beta_{xy} v_{xy}^{\orho}(\newtext{\tau}) 
    \, ,
    \\
    \label{eq:TotEntrProdSIJP}
    \sigma_\newtext{\tau} 
    &\coloneqq 
    \sum_{x < y} 
        \log \left( 
                \frac{\orho_x(\newtext{\tau}) Q_{xy}(\bar{M}_\newtext{\tau})}
                {\orho_y(\newtext{\tau}) Q_{yx}(\bar{M}_\newtext{\tau})} 
            \right) 
        v_{xy}^{\orho}(\newtext{\tau}) 
    \, .
\end{align}
\newtext{Over the backward logarithmic time scale, }\deletetext{Here} $j_\newtext{\tau}^{\orho}$ is the typical instantaneous current, such that \newtext{$\int_0^\infty e^{-\tau} j_\tau^{\orho} \, d\tau = \oj$} \deletetext{in the long-time limit}, and 
$\sigma_\newtext{\tau}$ is the instantaneous \textit{entropy production rate of the SIJP}\newtext{. The interpretation follows by}\deletetext{,} \newtext{viewing $\bar{M}_\newtext{\tau}$} \deletetext{can be thought of} as an ``internal'' adiabatic protocol\newtext{: since the empirical occupation measure evolves slowly, $Q(\bar{M}_\tau)$ changes quasi-statically, allowing the underlying SIJP to equilibrate at each frozen value of time and $\orho_\tau$ to be stationary for $Q(\bar{M}_\tau)$~\cite{Jarzynski1997,Seifert2018}. Instantaneous entropy-production rates have also appeared in non-Markovian Langevin settings, including active Ornstein--Uhlenbeck particles~\cite{Caprini2019} and generalized Langevin dynamics with memory~\cite{di-terlizzi2020a-thermodynamic}, where irreversibility is likewise quantified through path-weight/dynamical-cost comparisons with auxiliary processes.}

\newnewtext{Assuming that the observable \eqref{eq:GenCurrSIJP} satisfies a central limit theorem in the long-time limit, }\deletetext{E}\newnewtext{e}xpanding \er{eq:TURBoundRate} to second order around $\oj$ gives a bound to the quadratic fluctuations of the current 
\begin{equation}
    \label{eq:TURSIJPs}
    \frac{{\rm var}(j)}{\oj^2}
    \geq
    2
    \int_0^\infty 
        e^{-\newtext{\tau}} \frac{(j_\newtext{\tau}^{\orho} / \oj)^2}{\sigma_\newtext{\tau}} 
        \, d\newtext{\tau}
    \, ,
\end{equation}
which we refer to as SIJP-TUR, generalising the TUR of continuous-time Markov processes \cite{barato2015thermodynamic,gingrich2016dissipation}. As in the case of the TUR, \er{eq:TURSIJPs} shows that increasing precision for the current incurs a cost in terms of instantaneous entropy production. Since it is not possible to disentangle in the integral
the instantaneous entropy cost from the \deletetext{the} instantaneous current, we can interpret the instantaneous entropy production as an additional discount on the SIJP necessary to reach a certain level of precision. 

Note that both the SIJP-KUR \eqref{eq:KURSIJP} and SIJP-TUR \eqref{eq:TURSIJPs} reduce to the standard Markovian KUR and TUR in the limit when the SIJP becomes a Markov process. Note  also that the SIJP-TUR \eqref{eq:TURSIJPs} would hold in the case of {\em external} driving, as long as the protocol evolves as slowly as $A_t$ in a SIJP. 

\paragraph*{\bf \em Simple examples.} We now illustrate the general results above with two simple examples.  The first example is a two-state SIJP, $\S = \{0,1\}$, with transition rates $Q_{01} = 1 + e^{\alpha A_t}$, with $A_t = L_1(t)$, and $Q_{10} = 1$, with $\alpha > 0$. This describes a spin whose flip-up rate grows exponentially with its past occupation. There is a unique equilibrium state $\pi$, which for example for $\alpha = 2$ is  $\pi_0 = 0.12...$ (with $\pi_1 = 1 - \pi_0$), thus favouring occupation of the up state.

Figure~\ref{fig:all}(a) illustrates the level 2.5 result by showing the rate function at ``Level 2'' $I_2(\ell)$, obtained minimising \er{eq:RateFunct25} over the flux $\phi$. This exact result coincides with Monte Carlo simulations. We also show an upper bound obtained by assuming solutions of the variational problem to be time-independent \cite{donsker1975asymptotic}: for larger deviations $I_2(\ell)$ departs the Markov approximation
because larger fluctuations entail sharper variations in $\rho(\newtext{\tau})$ and $M_\newtext{\tau}$, see Fig.~\ref{fig:all}(b).

Figure~\ref{fig:all}(c) shows the exact rate function $I(b)$ for the flow observable $B_T = \Phi_{01}(T)$ obtained by contraction from \er{eq:RateFunct25}, and which coincides with the Monte Carlo simulations (whose range is limited to the values around the typical). We also show the SIJP-KUR upper bound from the r.h.s.\ of \er{eq:IbBoundSIJP}.

We illustrate the SIJP-TUR in a system with a non-zero stationary current, that of a particle on a three-state ring, $\S = \{0,1,2\}$, with a self-induced drag and cyclic transitions: $Q_{01}=Q_{12}=Q_{20}=e^{-\alpha L_1(t)}$ and $Q_{10}=Q_{21}=Q_{02}=1-e^{-\alpha L_1(t)}$. Dynamics slows down when state $1$ is visited frequently, and the more the particle lingers in one region, the harder it becomes to sustain a current. In Fig.~\ref{fig:all}(d) we compare the statistics for the particle current,  
$J_T = \sum_{x<y} (\Phi_{xy}(T) - \Phi_{yx}(T))$ from Monte Carlo simulations, with the SIJP-TUR bound for the rate function given in \er{eq:TURBoundRate}.

\paragraph*{\bf \em Conclusions.} Self-interacting processes offer a rich setting for uncovering universal principles of non-Markovian dynamics. \deletetext{Here}\newnewtext{In the regime where the relevant timescales are well separated and the dynamics is sufficiently mixing,} we have derived the exact dynamical large deviations at level 2.5, that is the full joint statistics of the empirical measure and fluxes, of self-interacting continuous-time processes.

For concreteness, we focused on SIJPs with generators dependent on empirical observables of configurations. Our results, however, are easily generalisable to self-interacting systems where the dynamics depend\deletetext{end}s also on the accumulated fluxes: if $A_t$ in \er{eq:EmpObs} also adds $t^{-1} g_{xy}$ for every jump between $x$ and $y$ in a trajectory, the corresponding level 2.5 LDs are 
also given by \ers{eq:RateFunct25}{eq:RareFlux}, with the change
$\sum_x f_x \rho_x \to 
\sum_x f_x \rho_x + \sum_{xy} g_{xy} \rho_x H_{xy}$
in the integrand on the r.h.s.\ of \er{eq:Ms}. See Ref.\ssp\cite{coghi2026proof} for a detailed derivation.

As is the case with Markovian systems, an important application of the level 2.5 LDs framework is in the derivation of bounds on the fluctuations (to all orders) of dynamical observables, which for quadratic fluctuations are the so-called thermodynamic and kinetic uncertainty relations.
Here, we obtained the corresponding versions for SIJPs of the 
thermodynamic \cite{barato2015thermodynamic,gingrich2016dissipation}
and kinetic 
\cite{garrahan2017simple,terlizzi2018kinetic}
uncertainty relations, the SIPJ-TUR and SIPJ-KUR. 
More refined and tighter bounds, such as a generalisation of the ``ultimate'' KUR \cite{macieszczak2024ultimate} (also known as the ``clock'' uncertainty relation \cite{prech2024optimal}) can be obtained by generalising the approach of \cite{macieszczak2024ultimate} to SIJPs, see \cite{coghi2026proof}.

\newtext{Looking ahead, t}he present framework could, in principle, also be adapted to derive bounds for first-passage times \cite{garrahan2017simple,gingrich2017fundamental} beyond the Markovian regime. \newtext{More complex illustrative examples and applications--such as} extension\newtext{s} to open quantum dynamics with memory--\deletetext{would also be interesting}\newtext{are important for clarifying the scope and implications of the theory. Another natural direction is to reweight past memory, for example by damping contributions from the distant past or restricting the memory to a finite time window, as would arise from rapidly evaporating pheromone trails in autochemotactic microorganisms~\cite{budrene1995dynamics,brenner1998physical}.}

\paragraph*{\bf \em Acknowledgments.}

F.C.\ is grateful to Tobias Grafke for help on clarifying numerical aspects behind the two-state examples. F.C.\ is supported by a Leverhulme Early Career Fellowship No.\ ECF-2025-482. We acknowledge support from the EPSRC Grants No.\ EP/V031201/1 and EP/T022140/1. A.B. supported in part by the NSF (DMS-2134107, DMS-2152577, DMS-2506010) and DARPA ARC Compass project HR0011-25-3-0240.

\paragraph*{\bf \em Appendix: notation and conventions.} 
\label{appendix}
We use upper case for random variables (or functions) and lower case for their realisations, while reserving $P$ for probability: $P(X=x)$ means the probability of the random variable $X$ taking value $x$, for which we often just write $P(x)$. 
We use a compact notation for multidimensional objects: given vector $\pi \coloneqq (\pi_x)_{x=1:d}$ and matrix $Q \coloneqq (Q_{xy})_{x,y=1:d}$, $\pi Q$ denotes their internal product, $(\pi Q)_y = \sum_{x=1}^d \pi_x Q_{xy}$, and use $\circ$ for Hadamard product, with $\pi \circ Q$ having elements $\pi_x Q_{xy}$. 
For scalar quantities we write the time dependence with a subscript, e.g.\ $A_t$, for compactness.
$\mathbf{1}_x(\cdot)$ is the indicator function, which gives one if the argument coincides with $x$ or zero otherwise. 
We use overbar to mark typical values, e.g., $\lim_{t \rightarrow \infty} A_t = \bar{a}$. 
The symbol $\asymp$ indicates equality up to sub-exponential factors in a large parameter, so that $P(\cdot) \asymp e^{-T I(\cdot)}$ is equivalent to $I(\cdot) = \lim_{T \to \infty} T^{-1} \log P(\cdot)$. 
We define the asymptotic variance of an (intensive) trajectory observable such as $B_T$ as ${\rm var}(b) \coloneqq \lim_{T \to \infty} T \, {\rm var}(B_T)$. 
We use the ``maths'' convention of probabilities being row vectors, stochastic generators having rows adding up to zero, and time-propagation being left to right, e.g.\ $\partial_t \mu = \mu Q$.

\bibliography{bibliography-04012026}
\bibliographystyle{apsrev4-2}

\end{document}